# Bloch-Siegert oscillations in the Rabi model with an amplitude-modulated driving field

A.P. Saiko [1], S.A. Markevich [1], R. Fedaruk [2]

[1] Scientific-Practical Material Research Centre, Belarus National Academy of Sciences, 19 P.Brovka str., Minsk 220072 Belarus

[2] Institute of Physics, University of Szczecin, 15 Wielkopolska str., 70-451, Szczecin, Poland

E-mail: saiko@physics.by

**Abstract**

We study the coherent dynamics of a qubit excited by an amplitude-modulated electromagnetic field under the Rabi resonance when the frequency of the low-frequency modulation field matches the Rabi frequency in the high-frequency field. Due to destructive interference of multiple photon processes at the ultrastrong coupling between the qubit and the low-frequency driving field, Rabi oscillations result exclusively from the Bloch-Siegert effect. It is directly observed in the time-resolved coherent dynamics as the Bloch-Siegert oscillation. In this case, triplets in Fourier spectra of the coherent response are transformed into doublets with the splitting between the lines equal to twice the Bloch-Siegert shift. These unusual properties are demonstrated in conditions of experiments with a nitrogen vacancy center in diamond.

## 1. Introduction

The coherent dynamics of resonant interaction between a classical light field and a two-level system (qubit) is described by the semiclassical Rabi model [1]. This dynamics has widely been studied for various quantum objects, including nuclear and electronic spins [2, 3], natural atoms [4, 5], artificial atoms such as quantum dots [6] and superconducting qubits [7]. The quantum Rabi model [8, 9], where the light field is quantized, is used in various areas of quantum physics and, in particular, extremely important for quantum information processing [3].

In the weak-coupling regime, when the coupling is significantly smaller than the qubit transition frequency and field mode frequency, the dynamics of the qubit-field system is well described by the Rabi model within the rotating wave approximation (RWA), which neglects counter-rotating (non-RWA) off-resonant terms in the interaction Hamiltonian. However, in the ultrastrong coupling regime, when the coupling is comparable to the qubit's transition frequency and field mode frequency, the RWA breaks down, the non-RWA terms result in complex dynamics of the field-matter interaction (see, e.g. [10–13]) and make difficulties for its analytical description. The well-known influence of these terms is manifested in the appearance of the Bloch-Siegert effect [14]. As a result, the observed resonance frequency is shifted. This shift is usually negligible for optical transitions [4] but becomes significant for precision nuclear magnetic resonance (NMR) experiments [2], preventing accurate observation of the resonance frequencies. In the dispersive regime this shift sometimes is referred to as dynamical Stark shift [15].

Since the ultrastrong coupling regime has new perspectives not only in quantum information processing, but also in materials science, quantum chemistry, and quantum photonics [16], there is growing importance of the Bloch-Siegert effect. In this regime, significant contributions of the non-RWA terms hinder coherent control of qubits due to complex (multifrequency) Rabi oscillations [17].

The Bloch-Siegert effect appears important also for experiments with doubly dressed states of qubits driven by strong bichromatic field. In particular, the second low-frequency field with the frequency closed to the Rabi frequency in the first high-frequency electromagnetic field excites effectively transitions between the dressed states of the qubit. This so-called Rabi resonance has been observed in the optical range [18, 19] as well as in electron paramagnetic resonance [20–22] and NMR [23–25]. The Rabi oscillations between doubly dressed states have also been studied [19–22, 25]. Their frequency is determined by the amplitude of the low-frequency driving field. The bichromatic control of Rabi oscillations between doubly dressed states and prolongation of their coherence can find applications in quantum information processing [26, 27] and open the possibility for the direct and sensitive detection of weak radio-frequency magnetic fields [28]. Since the coupling between the qubit and the low-frequency field can easily be obtained comparable to the Rabi frequency in the high-frequency field, under such strong driving the Bloch-Siegert effect becomes significant [21, 22, 29]. In these time-resolved experiments, the Rabi resonance was realized when the high-frequency field was in resonance with the qubit transition. In this case the Bloch-Siegert effect was manifested in the shift of the Rabi frequency between doubly dressed states. Additional multiphoton resonances occur at the subharmonics of the Rabi frequency [18, 25, 30].

Recently, at multiphoton Rabi resonances so-called Floquet Raman transitions have been observed in the solid-state spin system of NV center in diamond driven by the microwave field with its low-frequency amplitude modulation [31]. It was shown [32] that the ultrastrong regime is reached in the experiment resulting in the significant Bloch-Siegert shift of the Rabi frequency of Raman transitions. More recently, at the stronger light-matter coupling, unexplored behaviors of Rabi oscillations for the second-order Raman transition was considered and propose the method for direct observation of the Bloch-Siegert oscillation [33]. Filtering the Bloch-Siegert oscillation from multiphoton coherent dynamics of a ultrastrongly driven qubit remains unstudied for other techniques of time-resolved coherent spectroscopy.

In the present paper, we demonstrate possibility for the conventional Rabi resonance in the qubit's states excited by the amplitude-modulated microwave field to observe the Bloch-Siegert oscillation separately from other oscillating processes in the coherent dynamics.

## 2. Coherent dynamics of the qubit in an amplitude-modulated driving field

Quantum transitions between states of a spin qubit are excited by an amplitude-modulated microwave field $V(t) = \Delta_x \cos(\omega_d t) + 2A\cos(\omega_d t)\sin(\omega t)$, where $\cos(\omega_d t)$ and $\sin(\omega t)$ describes the high- and low-frequency components of the field with the frequencies $\omega_d$ and $\omega$, respectively, and the amplitudes of these components $\Delta_x$, $A \ll \omega_d$ [31]. The Hamiltonian of the qubit at such driving can be written as $H_{lab} = \frac{\Delta E}{2}\sigma^z + \Delta_x \cos(\omega_d t)\sigma^x + 2A\cos(\omega_d t)\sin(\omega t)\sigma^x$, where $\Delta E$ is the transition energy between the ground and excited levels; $\sigma^z$ and $\sigma^z$ are Pauli operators. In the frame rotating with the driving field frequency $\omega_d$ and in the RWA for this field (since $\omega, \Delta_x, A \ll \omega_d$), the Hamiltonian is

$$H = \frac{\Delta_z}{2}\sigma^z + \frac{\Delta_x}{2}\sigma^x + A\sin(\omega t)\sigma^x, \qquad (1)$$

where $\Delta_z = \Delta E - \omega_d$. The dynamics of the system is described by the Liouville equation for the density matrix $\rho$: $i\partial\rho/\partial t = [H, \rho$ (in the following we take $\hbar = 1$). Rotating the frame around the $y$ axis by angle of $\theta$ ($\rho \to \rho_1 = U_1^+ \rho U_1$, $U_1 = e^{-i\theta\sigma^y/2}$, and $\sigma^y = (\sigma^+ - \sigma^-)/i$), we can write the same equation with the Hamiltonian $H_1 = U_1^+ H U_1 = \frac{\omega_0}{2}\sigma^z + A\cos\theta\sin(\omega t)\sigma^x + A\sin\theta\sin(\omega t)\sigma^z$, where $\omega_0 = \sqrt{\Delta_z^2 + \Delta_x^2}$, $\sin\theta = \Delta_x/\omega_0$, and $\cos\theta = \Delta_z/\omega_0$. After the second canonical transformation $\rho_1 \to \rho_2 = U_2^+ \rho_1 U_2$ with $U_2 = \exp\left\{-i\left[\omega_0 t - \frac{2A\sin\theta}{\omega}\cos(\omega t)\right]\frac{\sigma^z}{2}\right\}$, we obtain the Liouville equation for $\rho_2$ with the Hamiltonian

$$H_2 = U_2^+ H_1 U_2 - iU_2^+ \frac{\partial U_2}{\partial t} = \frac{A}{2i}\cos\theta\left[\sigma^+ \sum_{n=-\infty}^{\infty} J_n(a)e^{-in\pi/2}\left(e^{i(n+1)\omega t} - e^{i(n-1)\omega t}\right)e^{i\omega_0 t} + H.c.\right], \qquad (2)$$

where $J_n(a)$ is the Bessel function of the first kind and $a = 2A\sin\theta/\omega$.

We consider only the conventional Rabi resonance realized at $\omega_0 = \omega$. This resonance is well observed and has the largest Rabi frequency among others multiphoton Rabi resonances [32]. The Hamiltonian $H_2$ contains an infinite sum of oscillating harmonics with the frequencies which are integer multiples of the frequency $\omega$. There are no oscillations for $n = 0$ and $n = -2$. Therefore, the terms of the sum with these $n$ give the largest contribution and correspond to the RWA. At the strong coupling condition $0.1 < A\cos\theta/\omega < 1$ the other oscillating terms are significant. Their contribution can be taken into account using the Bogoliubov averaging method [34] for constructing time-independent effective Hamiltonian in the framework of the non-secular perturbation theory. The averaging procedure up to the second order in $A\cos\theta/\omega$ (see [32]) gives the following effective Hamiltonian: $H_2 \to H_{eff} = H_2^{(1)} + H_2^{(2)}$, where

$$H_2^{(1)} = <H_2(t)>, \quad H_2^{(2)} = \frac{i}{2}<[\int^t d\tau(H_2(\tau) - <H_2(\tau)>), H_2(t)]>. \qquad (3)$$

Here the symbol $\langle...\rangle$ denotes time averaging over rapid oscillations of the type $\exp(\pm im\omega t)$ given by $\langle O(t)\rangle = \frac{\omega}{2\pi}\int_0^{2\pi/\omega} O(t)dt$. The upper limit $t$ of the indefinite integral indicates the variable on which the result of the integration depends, and square brackets denote the commutation operation.

As a result, the effective Hamiltonian can be written as $H_{eff} = (\omega^{BS}/2)\sigma^z + (\Omega/2)(\sigma^+ + \sigma^-)$ with

$$\Omega = 2\frac{J_1(a)}{a}A\cos\theta, \quad \omega^{BS} = \frac{A^2\cos^2\theta}{2\omega}\left\{\sum_{n\neq -2}\frac{J_n^2 + J_n J_{n+2}}{n+2} + \sum_{n\neq 0}\frac{J_n^2 + J_n J_{n-2}}{n}\right\}, \qquad (4)$$

where $\Omega$ is the Rabi frequency for the quantum transitions, when the Rabi resonance condition $\omega_0 = \omega$ is fulfilled, and $\omega^{BS}$ is the Bloch-Siegert frequency shift caused by the non-resonant rapidly oscillating non-RWA terms. The Bessel function $J_1(a)$ in Eq.(4) for $\Omega$ appears due to virtual multiphoton transitions, in which the number of absorbed (emitted) photons exceeds by 1 the number of emitted (absorbed) photons. In the equation for $\omega^{BS}$ we omit the argument $a$ of the Bessel functions.

Now we consider the experimental situation with a NV center when the microwave field excites transitions between the spin sublevels $|0\rangle$ and $|-1\rangle$ of this center, while the level $|+1\rangle$ is far detuned [31]. We assume that the two-level spin system is initially in the ground state $|0\rangle$. At the Rabi resonance, the probability to find the system in some moment again in the ground state $P_{|0\rangle}(t)$ is:

$$P_{|0\rangle}(t) = \frac{1}{2}(1 + \cos^2\theta - 2c_1) + e\sin[\omega t - a\cos(\omega t)] + c\cos(\Omega^* t - \varphi_c) +$$
$$+ \frac{1}{2}\sin\theta\left\{b\cos(\Omega^* t - \varphi_b)\cos[\omega t - a\cos(\omega t)] + \frac{1}{2}d\cos(\Omega^* t - \varphi_d)\sin[\omega t - a\cos(\omega t)]\right\},\quad(5)$$

where the non-RWA Rabi frequency $\Omega^* = \sqrt{\Omega^2 + (\omega^{BS})^2}$ takes into account the Bloch-Siegert shift and the following definitions and notations are used: $c = (c_1^2 + c_2^2)^{1/2}$, $b = (b_1^2 + b_2^2)^{1/2}$, $d = (d_1^2 + d_2^2)^{1/2}$; $\cos\varphi_c = c_1/c$, $\cos\varphi_b = b_1/b$, $\cos\varphi_d = d_1/d$;

$$c_1 = \frac{1}{2}\left(\frac{\Omega}{\Omega^*}\right)^2\cos^2\theta - \frac{\omega^{BS}\Omega}{4\Omega^{*2}}\sin 2\theta\sin a, \quad c_2 = -\frac{\Omega}{4\Omega^*}\sin 2\theta\cos a,$$

$$e = -\left(\frac{\Omega}{\Omega^*}\right)^2\sin\theta\sin a - \frac{\Omega\omega^{BS}}{\Omega^{*2}}\cos\theta, \quad b_1 = \sin\theta\cos a, \quad b_2 = \frac{\Omega}{\Omega^*}\cos\theta - \frac{\omega^{BS}}{\Omega^*}\sin\theta\sin a$$

$$d_1 = \frac{\Omega\omega^{BS}}{\Omega^{*2}}\cos\theta - \left(\frac{\omega^{BS}}{\Omega^*}\right)^2\sin\theta\sin a, \quad d_2 = -\frac{\omega^{BS}}{\Omega^*}\sin\theta\cos a.$$

## 3. Time and spectral manifestations of the Bloch-Siegert effect

Figure 1 shows the dependences of the RWA Rabi frequency $\Omega$, the non-RWA Rabi frequency $\Omega^*$ and the Bloch-Siegert shift $\omega^{BS}$ on the normalized driving strength $a = 2A\sin\theta/\omega$. When approaching a value of $A^*/\omega$, where $A^*/\omega = a^*/2\sin\theta$ and $a^*$ is the first root of the equation $J_1(a) = 0$ in Eq. (4) for $\Omega$, the RWA Rabi frequency $\Omega$ tends to zero due to destructive interference of multiple photon processes, and the non-RWA Rabi frequency $\Omega^*$ reaches its minimum, i.e. $\omega^{BS}$. This first minimum is at the first zero crossing ($A/\omega = 2.00$) of the Bessel function $J_1(a)$. The frequencies $\Omega$, $\Omega^*$ and $\omega^{BS}$ near the first minimum ($\Delta A/\omega \equiv (A - A^*)/\omega = 0$) show in more detail in Fig. 1 (b). The non-RWA Rabi frequency reaches its minima at all values of $A/\omega$ corresponding the zero crossing of the Bessel function.

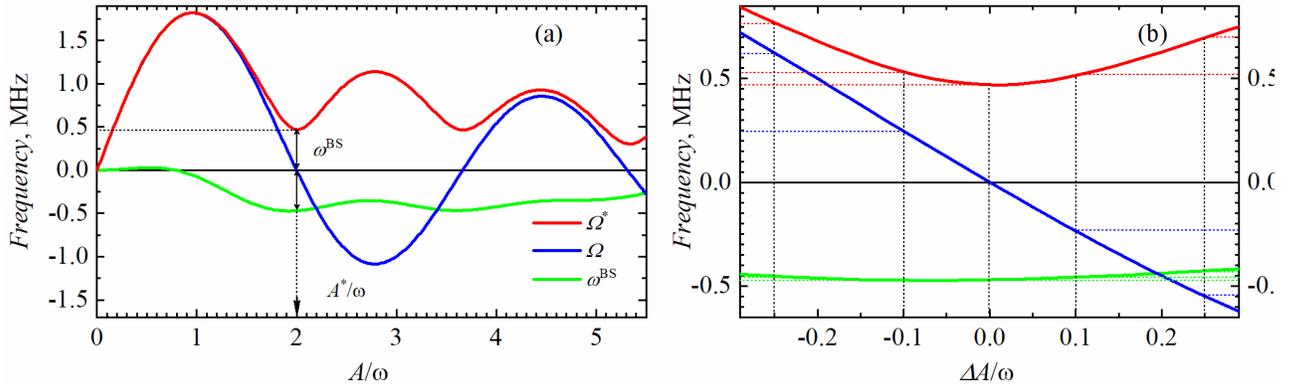

Fig 1. (a) The dependence of the RWA Rabi frequency $\Omega$, the non-RWA Rabi frequency $\Omega^*$, and the Bloch-Siegert shift $\omega^{BS}$ on the normalized amplitude of the low-frequency driving field at $\omega/2\pi = 10.44$ MHz, $\Delta x/2\pi = 10$ MHz, and $\Delta z/2\pi = 3$ MHz. (b) The frequencies presented in (a) near $A^*/\omega$ show in more detail. This plot is useful to obtain the values of these frequencies for $\Delta A/\omega$ used in the following figures.

To demonstrate the possibility of direct observation of the Bloch-Siegert effect, we use Eq. (5) for graphical illustration of the coherent dynamics of the qubit at the Rabi resonance for the parameters of the driving field which can be realized in experimental studies similar to those for NV center in diamond [31]. The first term of Eq. (5) for $P_{|0\rangle}(t)$ is time-independent. This term as well as the values of $A, e, b$ and $d$ are a function of the physical parameters of the quantum system and the driving high- and low-frequency fields. Using decomposition by a series of the Bessel functions, one can see that the second term of Eq. (5) describes rapid oscillations at the "carrier" frequencies $n\omega$, where $n$ is integer. The third term presents slow oscillations at the non-RWA Rabi frequency $\Omega^*$. The fourth and fifth terms describe oscillations at the "carrier" frequencies $n\omega$ with amplitudes modulated by the Rabi oscillations at the frequency $\Omega^*$. Fig. 2 illustrates the evolution of $P_{|0\rangle}(t)$ near and at the first minimum of the non-RWA Rabi frequency $\Omega^*$.

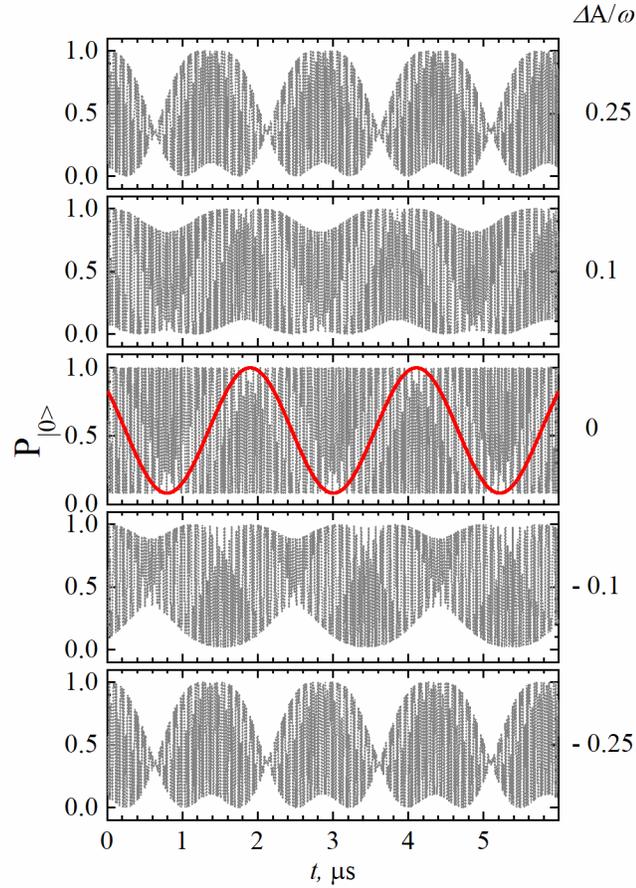

Fig 2. The state population of the spin level $|0\rangle$ at the Rabi resonance as a function of the evolution time at $\omega/2\pi = 10.44$ MHz, $\Delta x/2\pi = 10$ MHz, and $\Delta z/2\pi = 3$ MHz. The strength of the low-frequency driving field is $A = A^* + \Delta A$, where $A^*/\omega = 2.00$ and $\Delta A/\omega = 0.25, 0.1, 0, -0.1, -0.25$. The red line shows the Bloch-Siegert oscillation.

At the value of $A^*/\omega$ the RWA Rabi frequency $\Omega = 0$, the non-RWA Rabi frequency $\Omega^* = \omega^{BS}$, and $P_{|0\rangle}(t)$ can be written as

$$P_{|0\rangle}(t;\Omega=0) = \frac{1}{2}(1+\cos^2\theta) + \frac{1}{2}\sin^2\theta \cos\left[(\omega+\omega^{BS})t - a^*\cos(\omega t) + a^*\right]. \tag{6}$$

Figure 2 demonstrates that at $\Delta A/\omega = 0$ the amplitude modulation vanishes and the oscillations with the constant amplitude and periodically changing frequency occur. The disappearance of the amplitude modulation in the evolution of the ground state population is the evidence that the RWA Rabi frequency vanishes and the non-RWA Rabi frequency becomes equal to the Bloch-Siegert shift. In this case, the Bloch-Siegert oscillation is observed as the frequency modulation of the coherent response and is presented in Fig. 2 by the red line. The used parameters correspond to the ultrastrong regime with the coupling constant $A\cos\theta/\omega \approx 0.57$.

The disappearance of the RWA Rabi frequency represents a kind of electromagnetically induced transparency. The two-level system may become transparent under its bichromatic driving by the high- and low-frequency field [35, 36]. This effect is based on the destructive interference of excited multiple photon processes. When the non-RWA is taken in to account for the low-frequency field, the full electromagnetically induced transparency cannot be realized, because the Bloch-Siegert oscillation remains if even the RWA Rabi frequency vanishes.

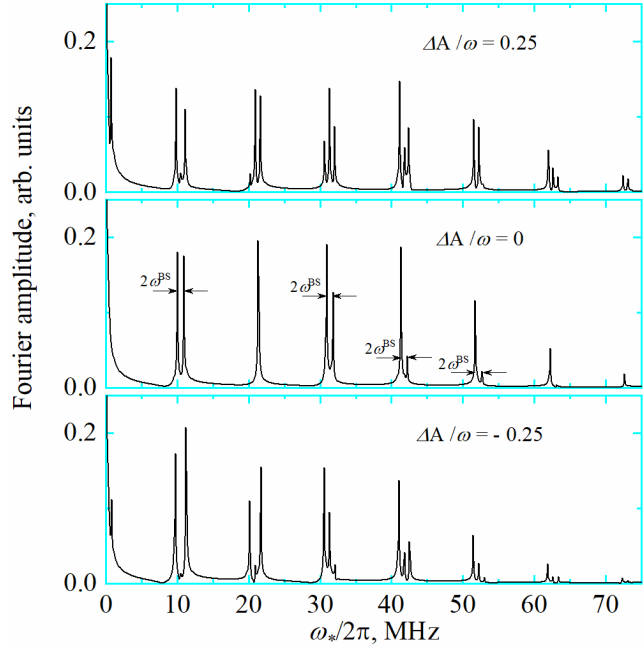

Fig 3. Fourier spectra for $\Delta A/\omega = -0.25$, 0 and 0.25. The other parameters are the same as in Fig. 2.

The Fourier spectra of $P_{|0\rangle}(t)$, $F(\omega_*) = \int_0^\infty e^{-i\omega_* t} e^{-\gamma t} P_{|0\rangle}(t) dt$, are shown in Fig. 3 for two values of the phase of the modulation field. The decay rate $\gamma$ was introduced phenomenologically using its value corresponding to a coherence time of 4 $\mu$s [31]. The spectra consist of Lorentzian lines at zero frequency, at the non-RWA Rabi frequency $\Omega^*$ and series of triplets are observed at frequencies $n\omega$ and $n\omega \pm \Omega^*$ corresponding to the amplitude-modulated oscillations. When $a \to a^*$ (or $A \to A^*$), the RWA Rabi frequency $\Omega \to 0$ as well as the coefficients $c_1, c_2, e \to 0$ and only the coefficients $b_1, b_2, d_1, d_2$ have non-zero values. In this case the line, corresponding to the oscillations at the RWA Rabi frequency, vanishes. At $A = A^*$, the non-RWA Rabi frequency $\Omega^* = \omega^{BS}$, only two side lines at the frequencies $n\omega \pm \omega^{BS}$ are remained in triplets and each triplet is transformed into doublet. A splitting between the doublet lines becomes exactly equal to $2\omega^{BS}$. Note that the second triplet near

$\omega_*/2\pi = 20.9$ MHz degenerates in a singlet. Indeed, using decomposition expansion of Eq. (6) by a series of the Bessel functions, the time-dependent part of this equation can be written as $\frac{1}{2}\sum_{n=-\infty}^{n=\infty} J_n(a^*)\exp\{i[((n-1)\omega - \omega^{BS})t - a^* + n\pi/2]\} + c.c.$ It follows directly from this expression that the line with the higher frequency in the doublet ($2\omega - \omega^{BS}, 2\omega + \omega^{BS}$) at $n = -1$ and $n = 3$ vanishes, because its amplitude $J_{-1}(a^*) = 0$. It is an additional indication that the RWA Rabi frequency $\Omega$ becomes equal to zero and the conditions are realized when the double Bloch-Siegert shift $2\omega^{BS}$ can directly be determined from the splitting between the doublet lines.

## 4. Conclusion

We have studied the coherent oscillations excited by the amplitude-modulated microwave field in the two-level system at the Rabi resonance. It was shown that in the ultrastrong regime, when the coupling between the qubit and the modulation field exceeds the modulation frequency, the Rabi oscillations are significantly modified by the Bloch-Siegert effect due to multiphoton antiresonant interactions. For properly chosen parameters of the modulation field, the RWA Rabi frequency can become zero and the Bloch-Siegert oscillation with the frequency $\omega^{BS}$ is directly observed. At these parameters the amplitude-modulated oscillations in the evolution of the ground state population transform into the oscillations of the constant amplitude and periodically changing frequency. In the Fourier spectra of the coherent response triplets are transformed into doublets with the splitting between the doublet lines equal to $2\omega^{BS}$. We demonstrate this unique possibility for the direct measuring the Bloch-Siegert shift in the conditions of experiments with the NV center in diamond. The proposed direct observation of the Bloch-Siegert oscillations may be used as a new technique for studying quantum systems at bichromatic and multichromatic driving in the ultastrong regime. This technique is not limited by spin qubits and can also be realized in quantum optics using an amplitude-modulated light.

## Acknowledgements

The work was supported by State Program of Scientific Investigations "Physical material science, new materials and technologies", 2016-2020, and by Project 01-3-1137-2019/2023, JINR, Dubna.